\documentclass[preprint,review,12pt]{elsarticle}
\usepackage{graphicx}
\journal{Computational Materials Science}

\begin{document}
\begin{frontmatter}
\title{\emph{Ab initio} study of $In_xGa_{1-x}N$ - performance of the alchemical mixing approximation.}

\author[PWR]{P. Scharoch}
\author[INT]{M.J. Winiarski}
\author[PWR]{M.P. Polak}
\address[PWR]{Institute of Physics, Wroc\ law University of Technology, Wybrzeze Wyspianskiego 27, 50-370 Wroc\l aw, Poland}
\address[INT]{Institute of Low Temperature and Structure Research, Polish Academy of Sciences, Ok\'olna 2, 50-422 Wroc\l aw, Poland}

\begin{abstract}
The alchemical mixing approximation which is the \emph{ab initio} pseudopotential specific implementation of the virtual crystal approximation (VCA), offered in the ABINIT package, has been employed to study the wurtzite (WZ) and zinc blende (ZB)  $In_xGa_{1-x}N$ alloy from first principles.  The investigations were focused on structural properties (the  equilibrium geometries), elastic properties (elastic constants and their pressure derivatives), and on the band-gap. Owing to the ABINIT functionality of calculating the Hellmann-Feynmann stresses, the  elastic constants have been evaluated directly from the strain-stress relation. Values of all the quantities calculated for parent $InN$ and $GaN$ have been compared with the literature data and then evaluated as functions of composition $x$ on a dense, $0.05$ step, grid. Some results have been obtained which, to authors' knowledge, have not yet been reported  in the literature, like  composition dependent elastic constants in ZB structures or composition 
dependent pressure derivatives of elastic constants. The band-gap has been calculated within the MBJLDA approximation. Additionally, the band-gaps for pure $InN$ and $GaN$ have been calculated with the Wien2k code, for comparison purposes. The evaluated quantities have been compared with the available literature reporting supercell-based \emph{ab initio} calculations and on that basis conclusions concerning the performance of the alchemical mixing approach have been drawn. An overall agreement of the results  with the literature data is satisfactory.  A small deviation from linearity of the lattice parameters and some elastic constants has been found to be due to the lack of the local relaxation of the structure in the VCA. The big bowing of the band-gap, characteristic of the clustered structure, is also mainly due to the lack of the local relaxation in the VCA.
The method, when applied with caution, may serve as supplementary tool to other approaches in \emph{ab initio} studies of alloy systems.
\end{abstract}


\end{frontmatter}

\section{Introduction}
Semiconducting group III metal nitrides have been drawing an interest over the last  decades because of their potential applications in optoelectronics. The direct band gaps  starting from 0.65-0.69eV \cite{PhysStatSolB_229_R1,APL_83_4963,JApplPhys_94_4457} for $InN$, through 3.50-3.51eV \cite{JApplPhys_83_1429, JApplPhys_94_3675}  for $GaN$, up to 6.1eV for $AlN$ \cite{JApplPhys_94_3675}, together with their ability to form ternary and quaternary alloys, open an interesting  perspective  for tuning the band gap which is of crucial importance for optoelectronic applications. The structural and electronic properties of nitride alloys have already been  intensively studied, both experimentally and theoretically.
The idea of tuning the band-gap, although simple in principle, is connected with a variety of practical problems like lattice constants mismatch of parent compounds, thermodynamically determined phase segregation, the effect of band-gab bowing, the efficiency of radiative transitions etc.
The \emph{ab initio} investigations are of particular importance in the field owing to their predictive power. They provide a hint in which direction, technologically and experimentally, to proceed. A lot of \emph{ab initio} works have already been  reported, dealing with the structural, elastic, thermodynamical  and electronic properties, including bulk systems, thin layers, and interfaces. A popular, supercell (SC) approach in which an alloy is modeled by periodically repeated large cell containing a few primitive cells offers an opportunity to vary compositions and ionic configurations. For example in wurtzite (WZ) structure  a 32-atoms cell (8 primitive cells) contains 16 nitrogen and 16 group III metal atoms which for ternary alloy gives 16 possible compositions and  a  number of configurations at each \cite{PRB_80_075202}. In ZB structure and 8 primitive cells in a supercell this number is respectively reduced. A great advantage of the supercell approach is the possibility to study the effect of 
various atomic configurations on physical properties, in particular the extreme cases of the clustered and the uniform one. However, the configurational space is still significantly limited by the supercell size which, if too big, leads to unrealistic computation time. For this reason for example, studying the alloy thermodynamics from first principles becomes a challenging task requiring various approximations \cite{Liu2008, Teles2002, Grosse2001}. Moreover, a simulated alloy is never a random alloy, i.e. the microscopic configuration of atoms in a supercell is periodically repeated which has an effect on the electronic structure \cite{Zunger,SQS}.

In this paper we employ an approach which is called the  alchemical mixing approximation, following the nomenclature introduced by the authors of the ABINIT package  \cite{ABINIT1,ABINIT2}. This is the modern, \emph{ab initio} pseudopotential based, implementation of the old idea of the virtual crystal approximation (VCA) whose main advantage is that the alloy can be studied within a primitive cell, which significantly reduces the computational costs. In the cell, at a metal site, the norm-conserving pseudopotential which is constructed of two pseudopotentials and which represents the scattering properties of two metal atoms entering the alloy is placed, at a given proportion. Thus, the composition becomes a continuous (not a discrete, like in supercells) parameter. One of important shortcomings of the approximation is that the "alchemical" atom is always on the ideal position, which means that the lattice distortion caused by different sizes of atoms, very characteristic of alloy systems, is not represented 
here, and which is (as we discuss later) the main reason of the deviation of the results from those obtained within the SC approach.  Also, studying the thermodynamics is not possible within the approximation since the lattice dynamics would be very poorly represented (the "alchemical" atom would have to have an unphysical intermediate mass). The aim of this work was to study the performance of the approximation in various applications, to find its strong or weak points and possible reasons of deficiencies, believing that when applied with caution can provide a useful reference for experiment and for the other \emph{ab initio} studies. We have concentrated on the structural, elastic and electronic properties. The ground state calculations gave us the opportunity to evaluate the LDA band-gap within the MBJLDA approximation \cite{MBJLDA}. An overall agreement of the results  with the literature data has appeared very satisfactory.  A small deviation from linearity of the lattice parameters and some elastic 
constants, showing an intermediate behavior between the clustered and the uniform structure of the alloy, has been found to be due to the lack of the local relaxation of the structure in the VCA. The apparent big bowing of the band-gap, characteristic of the clustered structure, points at certain inconsistency in the behavior of the VCA, which is supposed to simulate rather a perfectly uniform medium. An argumentation is given according to which  this effect is also mainly due to the lack of the local relaxation in the VCA.

\section{Computational methods}

The alchemical mixing of pseudopotentials implemented in the ABINIT package has been employed to emulate the $In_xGa_{1-x}N$ alloy. The prototype of the idea is the virtual-crystal approximation (VCA), used to describe mixed crystals within empirical potential approach. In the approximation the main idea is to introduce an object (an ion, scattering center) whose properties would reflect the properties of two atoms simultaneously, at a given proportion. The VCA is simply a linear combination of two one-electron potentials describing pure crystals. The alchemical mixing of pseudopotentials implemented in the ABINIT package uses the following construction \cite{ABINITwww}: the local potentials are mixed in the proportion given by mixing coefficients,
the form factors of the non-local projectors are all preserved, and all considered to generate the alchemical potential, the scalar coefficients of the non-local projectors are multiplied by the proportion of the corresponding type of atom, the characteristic radius of the core charge is a linear combination of the characteristic radii of the core charges, the core charge function $f(r/rc)$ is a linear combination of the core charge functions. In all the linear combinations the mixing coefficients reflecting the proportion at which particular atoms enter the alloy are used. Norm conserving pseudopotentials  with the same valence electronic configuration must be used, like e.g. In and Ga. It would be impossible then to emulate e.g. the $In_xAl_{1-x}N$ with In $d$-electrons included.

The norm conserving pseudopotentials have been generated with the OPIUM package \cite{opium}. The Perdew-Zunger form \cite{Perdew-Zunger} of the local density approximation (LDA) for the exchange-correlation functional was employed in the scalar relativistic mode. The cut-off radii: $2.0$, $1.8$, and $1.4$ $Bohr$ were selected respectively for In ($4d:5s:5p$), Ga ($3d:4s:4p$), and N ($2s:2p$) pseudo-orbitals. The non-linear core-valence correction (NLCV) radii \cite{fuchs} were: $1.0$, $0.8$ and $0.5$ $Bohr$, for In, Ga, and N, respectively. Psedopotentials were optimized with the Rappe-Rabe-Kaxiras-Joannopoulos method \cite{RRKJ}.

All the calculations have been performed with the ABINIT package \cite{ABINIT1, ABINIT2}. The total energy values were converged with the accuracy  $\approx 1meV$ on the $8 \times 8 \times 8$ Monkhorst-Pack grid  \cite{siatkaMP} with standard shifts for ZB and WZ structures \cite{ABINITwww}. Since the pseudpotentials used were rather hard, the 90Ha energy cut-off for the plane-wave basis set was used. The full geometry optimization, i.e. the cell and the ionic positions, has been performed  with standard convergence criteria for forces and stresses \cite{ABINITwww}.

The elastic constants have been evaluated from the stress-strain relation (the direct method). For ZB structure two strains have been applied: a tensile strain $(\epsilon,0,0,0,0,0)$ and a shear strain $(0,0,0,\epsilon,0,0)$ (in the Voight, vector notation). For WZ structure one more tensile strain $(0,0,\epsilon,0,0,0)$  was needed. In each case the calculations were performed for a few values of $\epsilon$, in the range $(\pm0.05,\pm0.2)$. At every value of $\epsilon$ the ions have been relaxed to their equilibrium positions. The values of
the stress tensor from the ground state calculations (Hellmann-Feynman stresses) were used to calculate the $C'_{ij}(\epsilon)$ constants from the stress-strain relation. Obtained in that way $\epsilon$ dependent $C'_{ij}$ values have been extrapolated to $\epsilon=0$ giving the elastic constants at equilibrium state $C_{ij}$, corresponding to infinitesimal strains. The pressure derivatives of elastic constants have been calculated as directional coefficients of straight lines fitted to 3-points.  The values of elastic constants, necessary for that purpose,  have been evaluated at 3 hydrostatic pressures (not exceeding 5 GPa)  in the same way as described above, except the preliminary ground state calculations were performed to find the reference states of a crystal at given pressure targets.

The related quantities, like the bulk modulus and Poisson's ratio have been calculated within the Voight-Reus-Hill approximation \cite{VRH52} (according to \cite{Wu07}). First, the Reuss (lower) \cite{Reuss} and Voight (upper) \cite{Voight} bounds, for the bulk ($B$) and for the shear ($G$) modulus have been evaluated, corresponding to policrystalline values at uniform stress and uniform strain respectively. Thus, for the cubic phase we have:

\begin{equation}
\begin{array}{l l}
B_V=B_R=(C_{11}+2C_{12})/3\\
G_V=(C_{11}-C_{12}+3C_{44})/5\\
G_R=5(C_{11}-C_{12})C_{44}/[4C_{44}+3(C_{11}-C_{12})]
\end{array}
\end{equation}

and for the hexagonal phase:

\begin{equation}
\begin{array}{l l}
B_V=(1/9)[2(C_{11}+C_{12})+4C_{13}+C_{33}]\\
G_V=(1/30)(M+12C_{44}+12C_{66})\\
B_R=C^2/M\\
G_R=(5/2)[C^2C_{44}C_{66}]/[3B_VC_{44}C_{66}+C^2(C_{44}+C_{66})]\\
M=C_{11}+C_{12}+2C_{33}-4C_{13}\\
C^2=(C_{11}+C_{12})C_{33}-2C_{13}^2
\end{array}
\end{equation}

Then the Young modulus ($E$) and the Poisson's ratio ($\nu$) have been calculated from the average values of B and G, $M_H=(1/2)(M_R+M_V), M=B,G$ (Voight-Reus-Hill approximation):

\begin{equation}
E=9BG/(3B+G), \nu=(1/2)(3B-2G)/(3B+G)
\end{equation}

Additionally, the ratio of shear modulus to bulk modulus B/G has been calculated to estimate the brittle
or ductile behavior of the material. A high B/G ratio is associated with ductility, whereas
a low value corresponds to the brittle nature. The critical value which separates ductile and brittle material
is $1.75$. If $B/G>1.75$, the material tends to be ductile, otherwise, it behaves in a brittle manner \cite{Pugh}.

The biaxial relaxation coefficients have been calculated from the formulae: $R^{WZ}_c=2C_{13}/C_{33}$ for WZ and
$R^{ZB}_c=2C_{12}/C_{11}$ for ZB structure.

The LDA band-gap as a function of composition has been calculated within the MBJLDA approximation \cite{MBJLDA}. The $C_m$ parameter for the parent compounds
has been fitted so that the values of band-gap it produced matched the experimental ones from \cite{}. It was then interpolated linearly to become a function of composition $x$.

\section{Results and discussion}

\begin{figure}
\centerline{\includegraphics[scale=1, angle=-90]{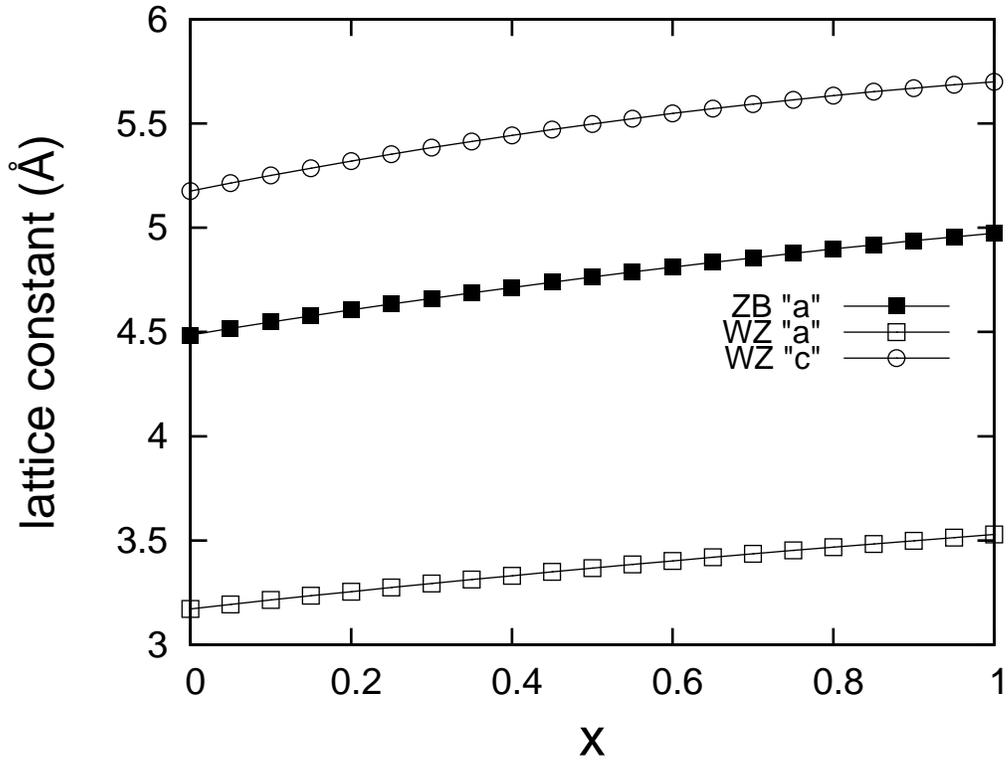}}
\caption{Equilibrium lattice constants for  $In_xGa_{1-x}N$ alloy; calculated values are marked with points}
\label{Facell}
\end{figure}

The \emph{ab inito} values  of equilibrium lattice parameters \textit{a} and \textit{c/a} ratio and internal parameter \textit{u} for parent $GaN$ and $InN$ compounds are presented in Tab.\ref{T_struc}. The quality of used pseudopotentials is confirmed by a good agreement with former results both experimental and theoretical. Fig.\ref{Facell} shows the composition dependence of the lattice constants of WZ and ZB structures which agree very well with independent supercell-based \emph{ab initio} calculations \cite{PRB_80_075202}. In Fig.\ref{Facell} it can be seen that the alloy lattice parameters for ZB and WZ structures change nearly linearly with the indium content \textit{x}, although a small deviation from linearity can be observed, especially for the $c$ parameter. The results reported in  \cite{PRB_80_075202} show that the linear composition dependence of the lattice constants (obeying Vegard's law) is characteristic of the uniform
configuration of indium whereas a small deviation from linearity of $c$ parameter appears in clustered configuration (Fig.1 in \cite{PRB_80_075202}).
The effect can be explained by the fact that when $InN$ component is gathered in clusters then it \textbf{}tends to keep its original lattice constant which is higher than that of $GaN$. Similarly, in the alchemical mixing approximation the atoms stay at their ideal lattice positions (do not relax), and the ''rigid`` contribution
of indium pseudopotential results in the bowing characteristic of the clustered case. Similar effect has been observed in $AlN_{1-x}P_x$ \cite{WPS2013} and
 $BN_{1-x}P_x$ alloys \cite{BNP}. According to our experience the effect of bowing in the VCA can be artificially suppressed by setting small orbital (hard) but big NLCC (soft) cut-off radii in the construction of pseudopotentials.

\begin{figure}
\centerline{\includegraphics[scale=1, angle=-90]{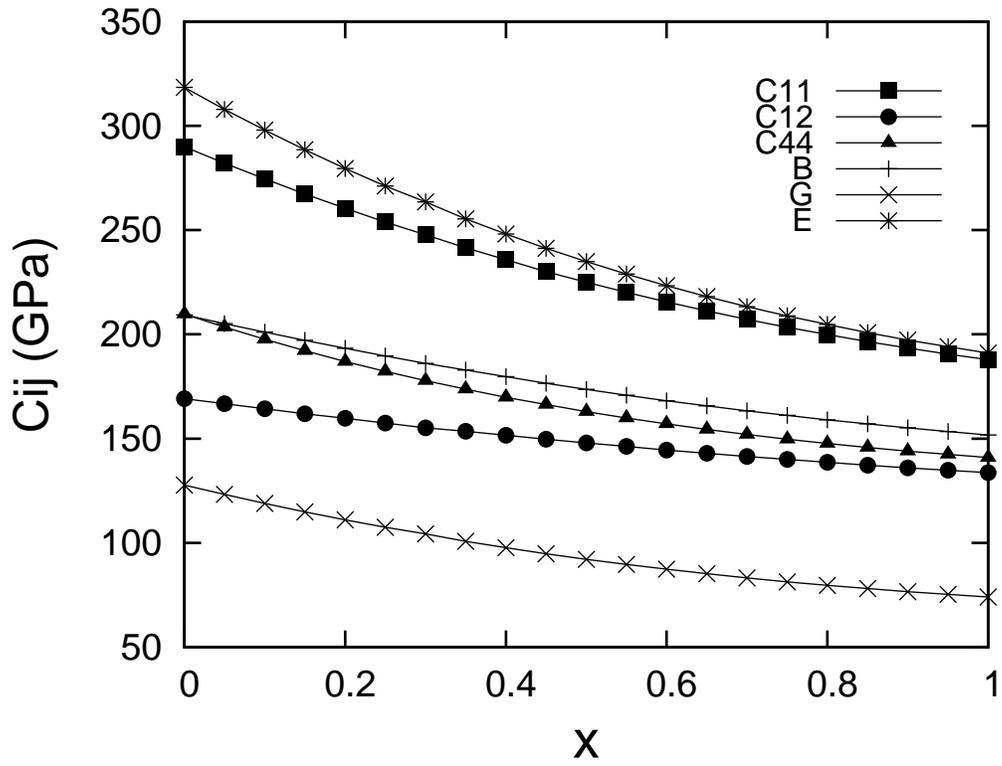}}
\caption{\emph{Ab initio} elastic constants ($C_{ij}$), bulk modulus ($B$), shear modulus  ($G$) and Young modulus ($E$) of  $In_xGa_{1-x}N$ alloy (ZB structure)}
\label{FCZB}
\end{figure}

\begin{figure}
\centerline{\includegraphics[scale=1, angle=-90]{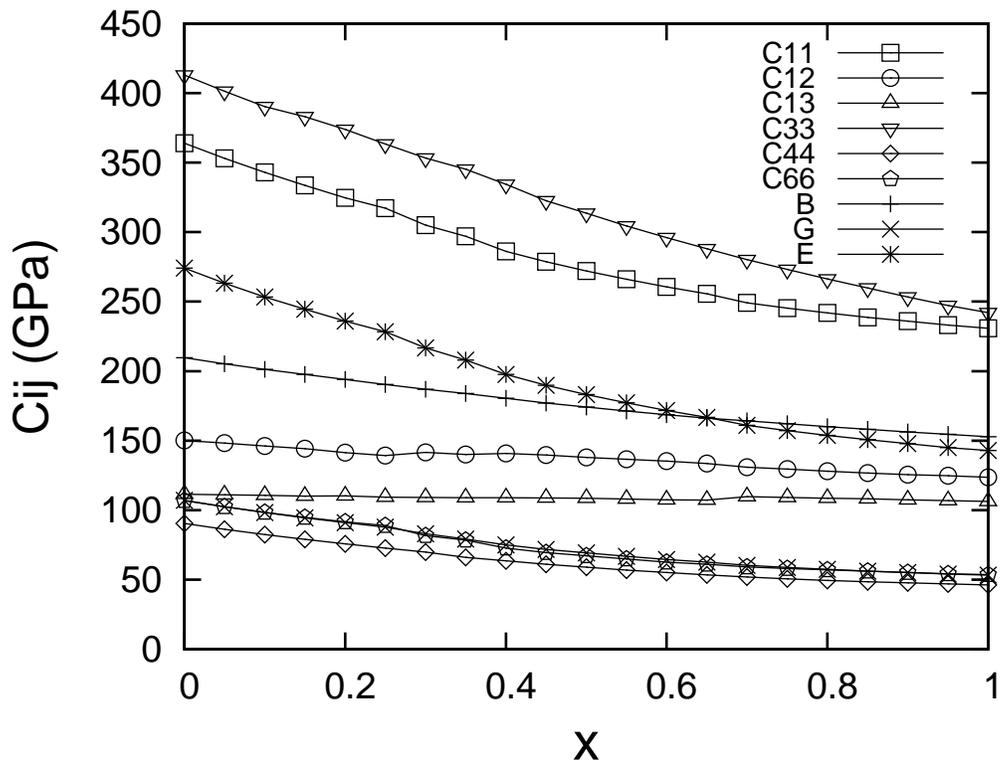}}
\caption{Elastic constants ($C_{ij}$), bulk modulus ($B$), shear modulus  ($G$) and Young modulus ($E$) of  $In_xGa_{1-x}N$ alloy (WZ structure)}
\label{FCWZ}
\end{figure}

In Tab.\ref{T_elast} the values of elastic constants calculated in this work  for parent $GaN$ and $InN$ are compared with the literature data, both theoretical and experimental. One can see that the present results fit well into the ranges of values reported earlier. One exception are the ZB $C_{44}$ constants whose values ($210GPa$ for $GaN$ and $141GPa$ for $InN$) are significantly larger than other reported values (by more than $30\%$). The problem has already appeared and has been discussed in the literature, namely, similar large values (respectively $206$ and $177GPa$) have been reported in \cite{KL96} and then corrected in \cite{KL97} to $142$ and $79GPa$. According to discussion in \cite{KL97} the discrepancy is due to simultaneous effect of semicore  $Ga$ 3d and  $In$ 4d states and high-lying conduction-band like $Ga$ 4d and $In$ 5d states on the valence band, which when poorly represented lead to high values of ZB $C_{44}$ constants. In the pseudopotential approximation used in this work, although 
the semicore  $Ga$ 3d and  $In$ 4d states are included, the high-lying conduction-band states are not represented sufficiently well. In Figs.\ref{FCZB} and \ref{FCWZ} the composition dependent elastic constants, bulk modulus, shear modulus and Young modulus are presented for ZB and WZ structures respectively. The results for WZ structure can be directly compared with the supercell based calculations reported in \cite{LG2011} and an excellent agreement can be observed. In the work \cite{LG2011} a distinction has been made between the case of the uniform and the clustered configuration of $In$ in $In_xGa_{1-x}N$. The particular elastic constants show either linear (Vegard's law) or sublinear  behavior with composition depending on the indium
configuration. We find our results to correspond neither to clustered nor to uniform case, although they are close to both, i.e. they represent an
intermediate (or mixed) state. To authors knowledge, there are no data for the composition dependent elastic constants in ZB structure (Fig.\ref{FCZB}) reported in the literature. In that case a small bowing (deviation from Vegard's law) is characteristic of all the dependencies.

\begin{figure}
\centerline{\includegraphics[scale=1, angle=-90]{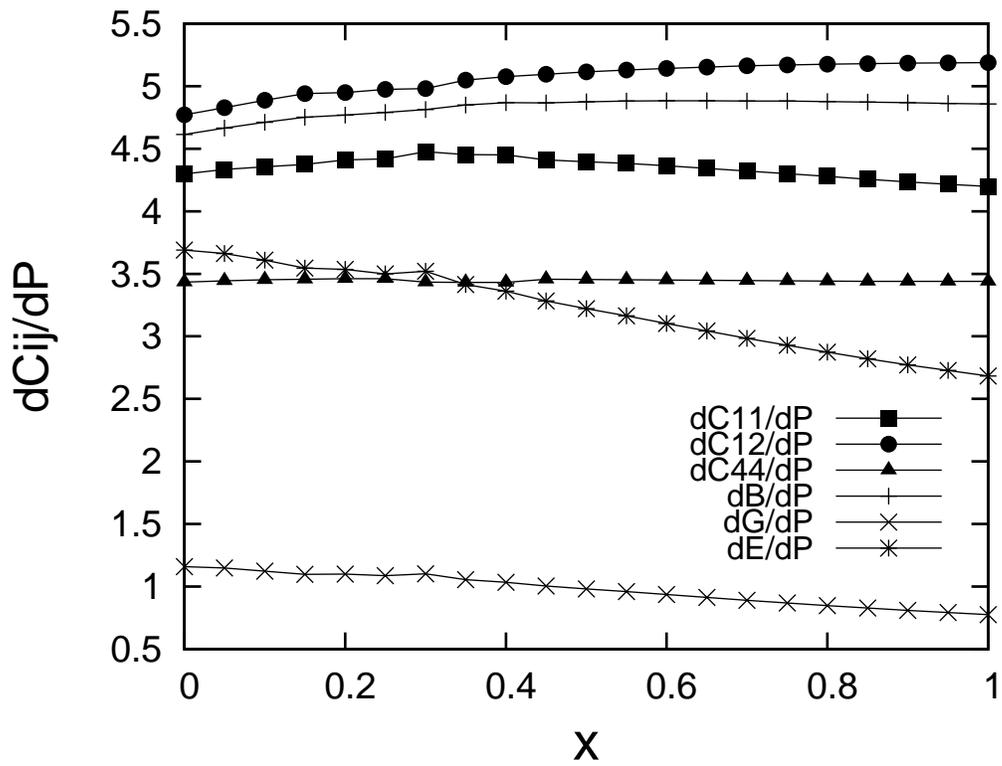}}
\caption{Pressure derivatives of elastic constants ($C_{ij}$), bulk modulus ($B$), shear modulus  ($G$) and Young modulus ($E$) of  $In_xGa_{1-x}N$ alloy (ZB structure)}
\label{FdCdPZB}
\end{figure}

\begin{figure}
\centerline{\includegraphics[scale=1, angle=-90]{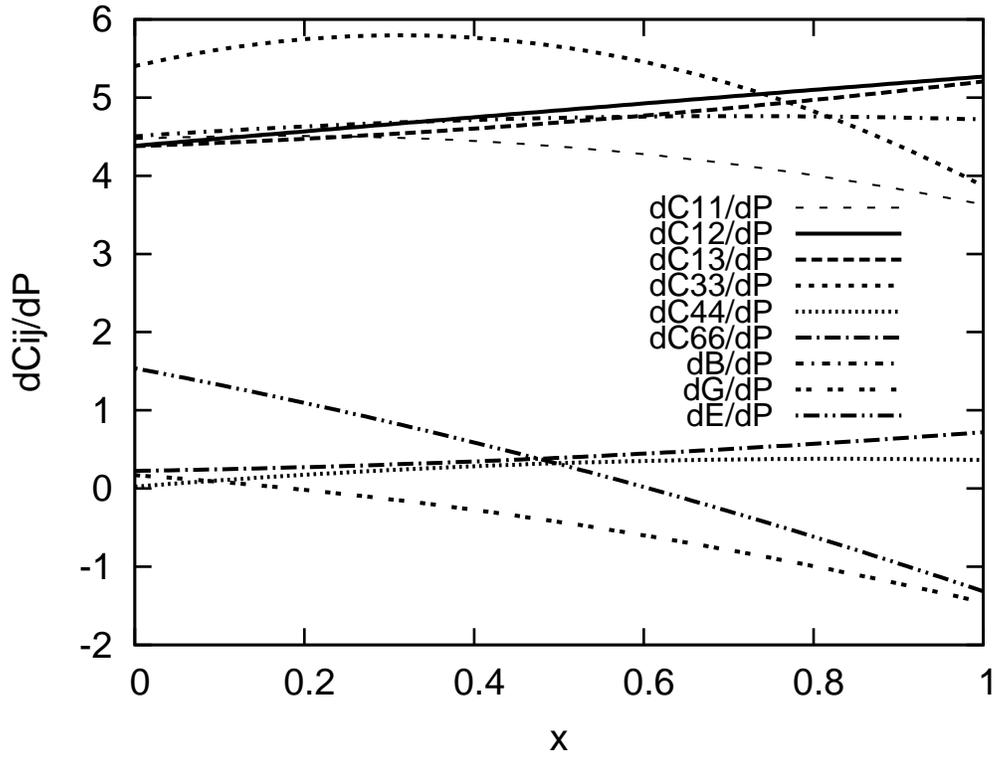}}
\caption{Pressure derivatives of elastic constants ($C_{ij}$), bulk modulus ($B$), shear modulus  ($G$) and Young modulus ($E$) of  $In_xGa_{1-x}N$ alloy (WZ structure); curves fitted to \emph{ab initio} data with standard deviation not exceeding 0.3.}
\label{FdCdPWZ}
\end{figure}

There are rather few works reporting pressure derivatives of elastic constants. Some reference data for parent $GaN$ and $InN$ are gathered in Tab.\ref{T_elastdP}, to compare them with the results of this work. The agreement of most of present results with the literature data is satisfactory, although some values ($dC_{12}/dP$, $dC_{13}/dP$) are higher (by $10-30\%$). A big difference can be seen in the case of $dC_{44}/dP$ which might be due to the reasons discussed in the previous paragraph. The original result of this work seem to be the composition dependent pressure derivatives of elastic constants, bulk modulus, Young modulus and shear modulus for ZB and WZ structures  which are shown in Figs.\ref{FdCdPZB} and \ref{FdCdPWZ}. In Fig.\ref{FdCdPZB} points represent the
\emph{ab initio}  data, whereas in Fig.\ref{FdCdPWZ}, to make the graph clearer, the second order polynomials have been fitted to \emph{ab initio} results with the standard deviation not exceeding $0.3$. The bowing (deviations from Vegard's) of majority of the quantities is rather small, except for  $dC_{33}/dP$ in WZ structure exhibiting an anomalously large bowing (a maximum of the pressure derivatives appears at $x=3.5$). In this work no second order term in the pressure
dependence of elastic constants has been evaluated. However, it is easy to estimate the expected corrections to the pressure derivatives which are due to the
second order term using data reported in \cite{SLL10,L07}. In most cases the correction is negative and small. For testing values of pressure applied in this work (up to $5GPa$) its value is between $0.1$ and $5$ percent.

\begin{figure}
\centerline{\includegraphics[scale=1, angle=-90]{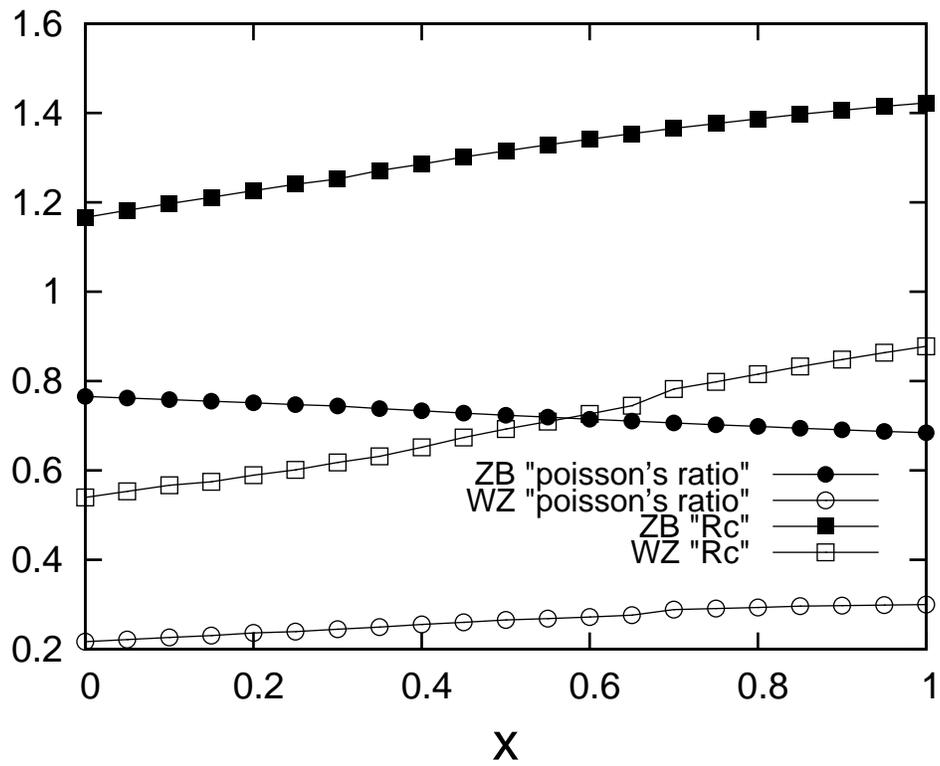}}
\caption{Poisson's ratio and biaxial relaxation coefficient of  $In_xGa_{1-x}N$ alloy (ZB and WZ structures)}
\label{FPR}
\end{figure}

In Fig.\ref{FPR} the composition dependent biaxial relaxation coefficient $R_c$ and Poisson's ratio $\nu$  are presented for ZB and WZ structures. The results can be compared with those reported in \cite{LG11}, where the parameters have been calculated for WZ structure \emph{ab initio} within the supercell approach and the effect of In distribution investigated. An excellent agreement of our results with those corresponding to the uniform distribution of In, for both parameters can be seen. Some reference values for parent $GaN$ and $InN$ in WZ structure can be found in \cite{KL94} and \cite{W97} where the reported (calculated) values  of the $R_c$ are respectively for GaN: $0.510,0.509$  and for InN: $0.940,0.821$, and are in a good agreement with present results.

\begin{figure}
\centerline{\includegraphics[scale=1, angle=-90]{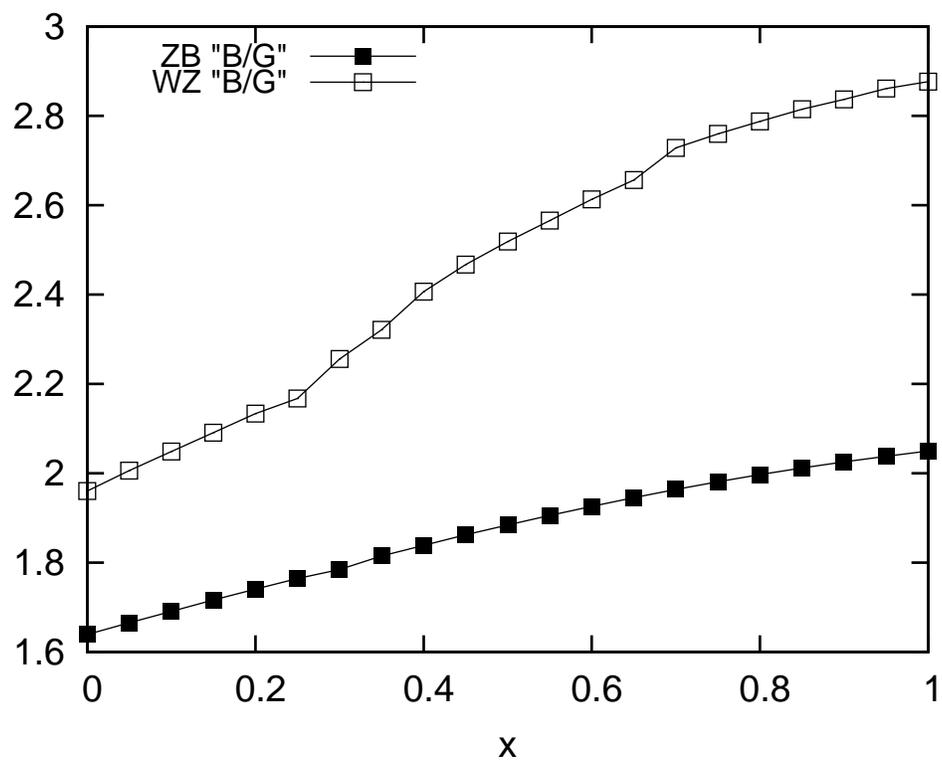}}
\caption{The $B/G$ of $In_xGa_{1-x}N$ alloy (ZB and WZ structures).}
\label{FBG}
\end{figure}

In Fig.\ref{FBG} the $B/G$ ratio is shown. Except for small range of $x\in(0.0,0.2)$ in ZB structure the material shows ductile character, according to the criterium presented in the previous chapter.

\begin{figure}
\centerline{\includegraphics[scale=1, angle=-90]{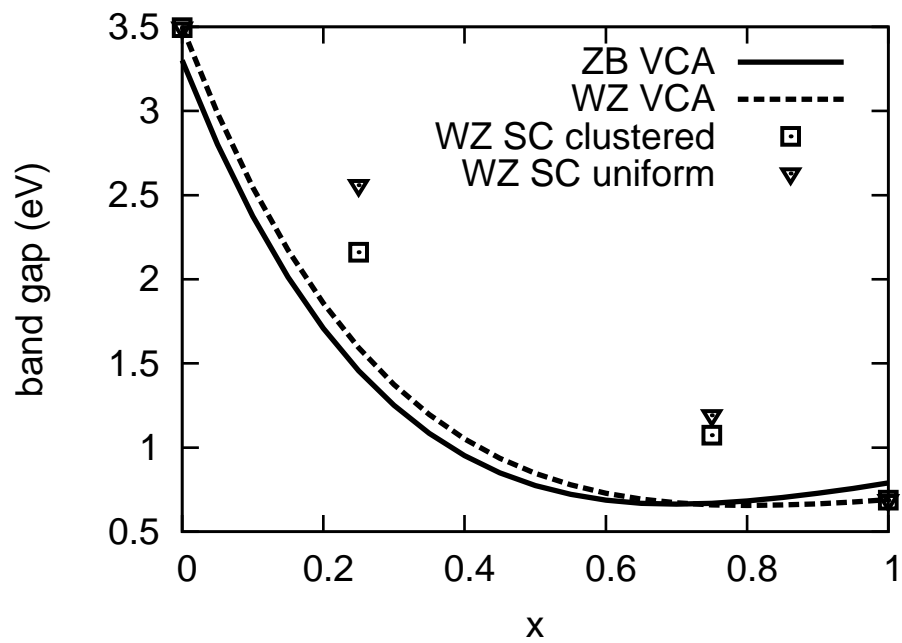}}
\caption{The band gap of $In_xGa_{1-x}N$ alloy (ZB and WZ structures) calculated with ABINIT (MBJLDA); the results of supercell calculations (WZ) are shown for comparison. }
\label{Fbandgap}
\end{figure}

Finally, in Tab.\ref{T_Eg} data for the band-gaps of parent $GaN$ and $InN$ obtained in this work and reported in the literature are collected. The values in this work has been obtained within MBJLDA approximation \cite{MBJLDA}, with the use of ABINIT and Wien2k codes. The values obtained with ABINIT coincide with
experimental ones owing to appropriate fitting of the $C_m$  parameter mentioned earlier. Its values are: $InN$(ZB) $1.505$ ($Eg=0.78eV$), $InN$(WZ) $1.36$ ($Eg=0.69eV$), $GaN$(ZB) $1.67$ ($Eg=3.30eV$), $GaN$(WZ) $1.63$ ($Eg=3.50eV$). For the alloy, its values are obtained form the linear interpolation. In Fig.\ref{Fbandgap} the $In_xGa_{1-x}N$ band-gap vs. composition is plotted, calculated within VCA. For comparison, the values of  Eg from SC calculations, for $x=0.25$ and $x=0.75$ and two In configurations (clustered and uniform), are given. The SC values of  the band-gaps are in good agreement with the values obtained in \cite{PRB_80_075202} and with experiment. However,  the band-gap  bowing obtained within VCA is bigger than even that in the SC clustered configuration, which somehow disagrees with expectation that the VCA should rather simulate a perfectly uniform alloy. For comparison, the calculations reported in \cite{CK11}, based on LMTO-CPA-MBJ for WZ $In_xGa_{1-x}N$ show smaller bowing of the 
band-gap (corresponding rather to the uniform configuration of In \cite{PRB_80_075202}) than in the present work. Bellow, an argumentation  according to which the anomalous bowing in VCA
 can be attributed to the fact that the "alchemical" atoms are always in ideal (unrelaxed) positions  and thus the distance between metal and nitrogen atoms is averaged, is given.

\section{An anomalous $In_xGa_{1-x}N$ band-gap bowing  in alchemical mixing approximation}

The admixture of $InN$ in $GaN$  leads to a lowering  of the band gap for two reasons: first, $InN$ has in nature much lower band-gap than $GaN$, and second, the involved expansion of the lattice constants leads to a lowering of the band gap in $GaN$. The latter effect appears also in $InN$ and is responsible for the difference in the band-gap between the uniform and the clustered configuration of In \cite{PRB_80_075202}, i.e. the bigger bowing for the clustered case is due to the locally expanded bonds between $In$ and $N$ atoms in the $InN$ cluster region. Thus, the band-gap appears to be very sensitive to bond lengths.  As mentioned above, in the VCA the bond lengths between the metal and the nitrogen atom are averaged, i.e. they are the same for partially contributing In and Ga. For example, at $x=0.25$, in SC uniform case the distances are:  $Ga$-$N$ $1.94 \AA$,  $In$-$N$  $2.12\AA$, whereas in VCA, metal-$N$ $2.01$.  The differences do not exceed $5\%$ but as will be shown below they are crucial for 
the band-gap behavior.

\begin{figure}
\centerline{\includegraphics[scale=0.5, angle=-90]{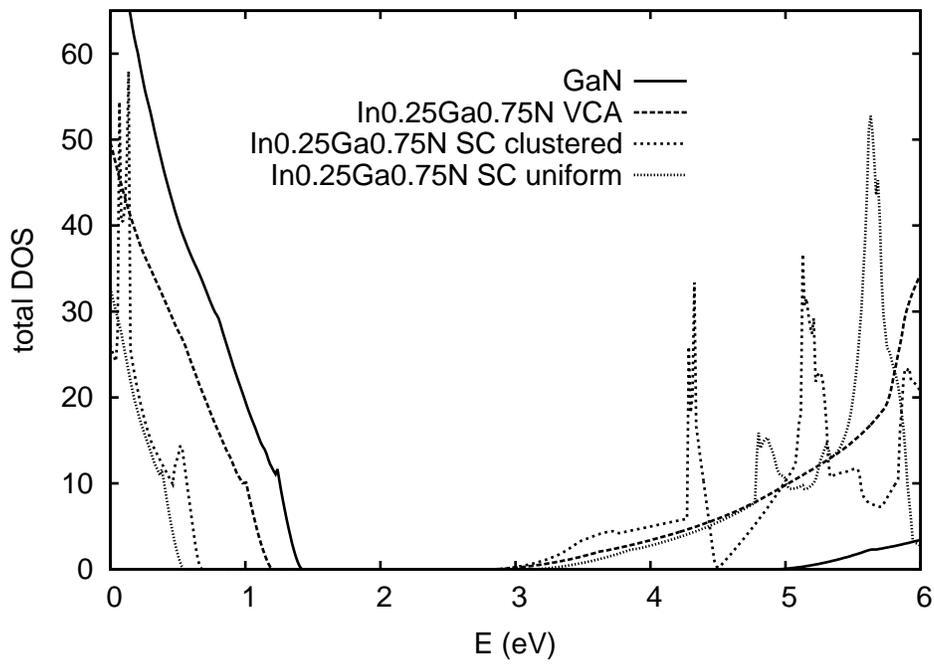}}
\caption{The effect of  $In$ doping ($x=0.25$) in $GaN$ on the total DOS  (the 32-atoms supercell DOS has been normalized to an elementary cell).}
\label{FDOScomp}
\end{figure}

\begin{figure}
\centerline{\includegraphics[scale=0.5, angle=-90]{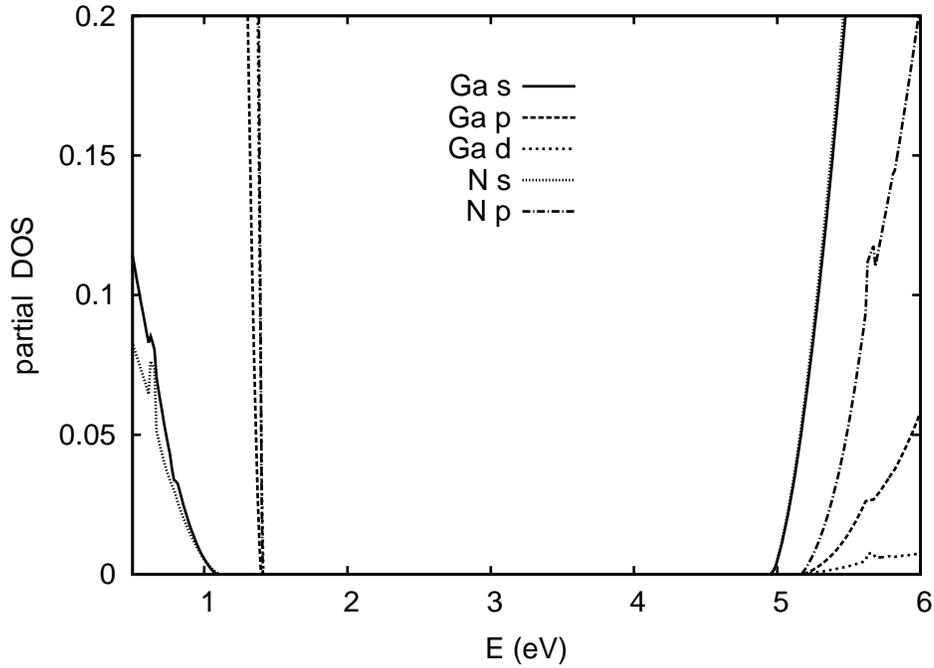}}
\caption{Near the band-gap fragment of the partial DOSes for $GaN$.}
\label{FDOSGaN}
\end{figure}

\begin{figure}
\centerline{\includegraphics[scale=0.5, angle=-90]{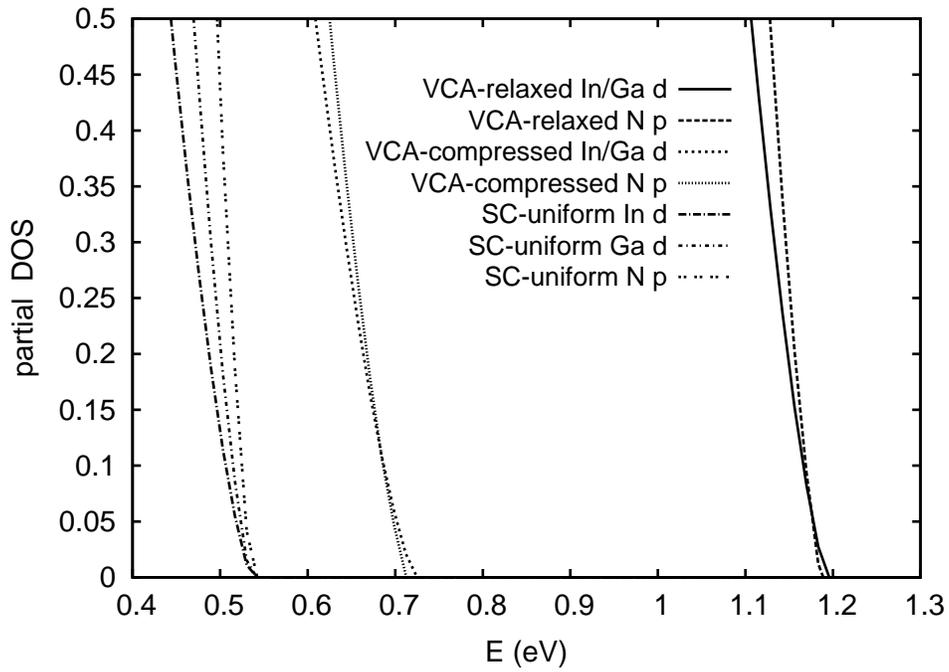}}
\caption{Partial DOSes near the VB for $In_xGa_{1-x}N$  at $x=0.25$, three cases: 1. VCA,  relaxed lattice, 2. VCA, lattice compressed by $5\%$, 3.  SC, a uniform In distribution (the bottom of the CB has been taken as the reference point).}
\label{FDOSSCalchcomp}
\end{figure}

In Fig.\ref{FDOScomp} the effect of $In$ doping  in $GaN$ is presented in terms of the total density of states. The large CB negative offset (the electron trapping case) appears to be the same in the VCA and the SC approaches, however, the VB offsets are very different. Thus, the direct reason for the large band-gap bowing in VCA is the wrong behavior of the VB with doping and can be understood by analyzing the  DOSes projected on the angular momentum eigenstates of chosen atoms (partial DOSes).
From Fig.\ref{FDOSGaN}, which shows a near the band-gap  fragment of the projected DOS for $GaN$,  it is clear that the main contribution to the bottom of CB comes from the metal $s$-state and  the $N$ $s$-state, whereas the top of the VB is formed mainly by the $N$ $p$-states and the metal $p$,$d$-states. The situation is the same in alloy simulated either within the VCA or SC.  Since the reason for the deep bowing in VCA is the VB behavior we will concentrate on that region now. In Fig.\ref{FDOSSCalchcomp}  the projected DOSes are plotted in the region of the VB  for three cases: 1. VCA,  relaxed lattice, 2. VCA, lattice compressed by $5\%$, 3.  SC, with the uniform  $In$ distribution.  A large shift of the VB towards the correct SC position can be seen for the compressed VCA lattice, which confirms the fact of high sensitivity of the bond lengths  on the band-gap and the presented above hypothesis of the bond length effect on the VCA band-gap.

It should be mentioned that the VCA band-gap behavior in alloy systems can be different in different systems. For example, in $AlN_xP_{1-x}$ the tendency is
opposite, i.e. the VCA shows smaller bowing than in SC based calculations \cite{WPS2013}. An analysis similar to that presented in this work should be done to explain this fact.  It's worth to add that some purely technical procedure, based on averaging over the transition energies near the transition point, can be applied within VCA approach to obtain the correct composition dependent band-gap of $InGaN$ alloy, as we have shown in [37]. Thus, in spite of the discussed difficulties, the alchemical pseudopotential method can be used
for band-gap calculations.

\section{Conclusions}

   The main objective of this work was to test the performance of the alchemical mixing of pseudopotentials approximation in theoretical \emph{ab initio} studies of structural, elastic and electronic properties of semiconductor alloys, on the example of $In_xGa_{1-x}N$. We find the results for various calculated parameters to be in an overall good or very good agreement with other \emph{ab initio} calculations, performed within the supercell approach, and with the experimental values. The behavior of the lattice parameters and the elastic constants together with the related quantities as a function of composition appears to be intermediate between the uniform and clustered structure of the alloy, whereas  the band-gap  would rather behave  like the alloy with clustered structure, which is, as discussed above, an artefact connected mainly with the VCA inherent feature of the lack of the local relaxation of atomic positions. This seems to be also the main reason for the discrepancy (although rather small) 
between other VCA and SC results. As an additional result of this work, some composition dependent quantities, such as composition dependent elastic constants and related quantities in ZB structures, or composition dependent pressure derivatives of elastic constants, have been calculated. Their values, to authors' knowledge, have not been reported previously. The obtained results lead to a conclusion that the \emph{ab initio} alchemical mixing approximation, if used with caution, can serve as a supplementary tool for semiconductor alloy studies.

\section{Acknowledgments}
Calculations have been carried out in Wroclaw Centre for Networking and Supercomputing.

\begin{table}\footnotesize

\caption{Lattice parameters of GaN and InN; \textit{a} in WZ and ZB structures; the \textit{c/a} ratio and internal parameter \textit{u} in WZ structure. }
\begin{flushleft}
\begin{tabular}{ccccc}
\hline
& &\textit{a} (\AA)&\textit{c/a}&\textit{u}\\
\hline
GaN-wz&this work&3.17&1.632&0.375\\
&other calc.&3.16$^d$,3.17$^c$&1.626$^d$,1.628$^c$&0.377$^{cd}$\\
&exp.&3.19$^a$&1.627$^a$&0.377$^a$\\
InN-wz&this work&3.53&1.615&0.377\\
&other calc.&3.50$^d$,3.52$^{gc}$&1.612$^c$,1.614$^g$,1.619$^d$&0.378$^d$,0.380$^{gc}$\\
&exp.&3.53$^f$&1.613$^f$&-\\
\hline
GaN-zb&this work&4.49&-&-\\
&other calc.&4.461$^j$,4.46$^k$,4.46$^m$&-&-\\
&exp.&4.5$^l$&-&-\\
InN-zb&this work&4.97&-&\\
&other calc.&4.932$^d$,4.95$^m$&-&-\\
&exp.&4.98$^l$&-&-\\
\hline
\end{tabular}\\

$^a$ Ref. \cite{SolStatComm_23_815}.\\
$^b$ Ref. \cite{PRB_49_14}.\\
$^c$ Ref. \cite{SemicondSciTech_18_850}.\\
$^d$ Ref. \cite{PRB_51_7866}.\\
$^e$ Ref. \cite{PRB_81_235206}.\\
$^f$ Ref. \cite{JApplPhys_46_3432}.\\
$^g$ Ref. \cite{JApplPhys_104_013704}.\\
$^h$ Ref. \cite{PRB_45_83}.\\
$^i$ Ref. \cite{InorgMater_15_1257}.\\
$^j$ Ref. \cite{KGM04}.\\
$^k$ Ref. \cite{KL94}.\\
$^l$ Ref. \cite{RS06}.\\
$^m$ Ref. \cite{CG94}.\\

\end{flushleft}
\label{T_struc}
\end{table}

\begin{table}\footnotesize
\caption{Elastic constants and bulk modulus of GaN and InN in WZ and ZB structures (in GPa).}
\begin{flushleft}

\begin{tabular}{ccccccccc}
\hline
System&Data from &$C_{11}$&$C_{12}$&$C_{13}$&$C_{33}$&$C_{44}$&$C_{66}$&$B$\\
\hline
&this work&290&169&-&-&210&-&209\\
&other calc.&293$^a$, 282$^b$&159$^a$, 159$^b$&-&-&155$^a$, 142$^b$&-&184$^n$,197.88$^o$\\
GaN-zb&&305$^f$,264$^g$&128$^f$,153$^g$&-&-&147$^f$,68$^g$&-&\\
&exp.&-&-&-&-&-&-&237$^n$,245$^n$,195$^n$\\
\hline
&this work&188&134&-&-&141&-&152\\
&other calc.&187$^a$, 182$^b$&125$^a$, 125$^b$&-&-&86$^a$, 79$^b$&-&137$^n$\\
InN-zb&&217$^f$,172$^g$&101$^f$,119$^g$&-&-&104$^f$,37$^g$&-&\\

&exp.&-&-&-&-&-&-&125.5$^n$\\
\hline
&this work&364&150&111&412&90&107&210\\

&other calc.&367$^a$,346$^b$&135$^a$,148$^b$&103$^a$,105$^b$&405$^a$,389$^b$&95$^a$,76$^b$&121$^f$&202$^a$,210$^m$\\

GaN-wz&&357$^f$,337$^h$&116$^f$,113$^h$&89$^f$,97$^h$&383$^f$,353$^h$&102$^f$,95$^h$&&\\

&exp.&390$^c$,374$^d$&145$^c$,106$^d$&106$^c$,70$^d$&398$^c$,379$^d$&105$^c$,101$^d$&&210$^c$,180$^d$\\
&&377$^i$,390$^j$&160$^i$,145$^j$&114$^i$,106$^j$&209$^i$,398$^j$,&81$^i$,105$^j$&&\\
\hline
&this work&231&124&106&242&46&54&154\\
&other calc.&223$^a$,220$^b$&115$^a$,120$^b$&92$^a$,91$^b$&224$^a$,249$^b$&48$^a$,36$^b$&82$^f$&141$^a$,152$^l$\\
InN-wz&&257$^f$,211$^h$&92$^f$,95$^h$&70$^f$,86$^h$&278$^f$,220$^h$&68$^f$,48$^h$&&\\

&exp.&190$^e$,223$^j$&104$^e$,115$^j$&121$^e$,92$^j$&182$^e$,224$^j$&10$^e$,48$^j$&&139$^e$,126$^k$\\
\hline
\end{tabular}\\

$^a$ Ref. \cite{W97}.\\
$^b$ Ref. \cite{KL97}.\\
$^c$ Ref. \cite{PG96}.\\
$^d$ Ref. \cite{TA96}.\\
$^e$ Ref. \cite{SS79}.\\
$^f$ Ref. \cite{MAG98}.\\
$^g$ Ref. \cite{SD91}.\\
$^h$ Ref. \cite{LM12}.\\
$^i$ Ref. \cite{M04}.\\
$^j$ recommended values Ref. \cite{VM03}.\\
$^k$ Ref. \cite{JApplPhys_104_013704}.\\
$^l$ Ref. \cite{SemicondSciTech_18_850}.\\
$^m$ Ref. \cite{PRB_81_235206}.\\
$^n$ according to Ref. \cite{CG94}.\\
$^o$ Ref. \cite{Y03}.\\
\end{flushleft}
\label{T_elast}
\end{table}

\begin{table}\footnotesize

\caption{Pressure derivatives of elastic constants and bulk moduli of GaN and InN in WZ and ZB structures.}
\centering
\begin{flushleft}
\begin{tabular}{ccccccccc}

\hline
System&Data from &$dC_{11}/dP$&$dC_{12}/dP$&$dC_{13}/dP$&$dC_{33}/dP$&$dC_{44}/dP$&$dC_{66}/dP$&$dB/dP$\\
\hline
GaN-zb&this work&4.3&4.8&0&0&3.4&0&4.6\\
&other calc.&3.88$^c$,3.64$^d$&3.33$^c$,4.87$^d$&0&0&1.02$^c$,-0.55$^d$&0&3.51$^c$,4.32$^d$\\
\hline
InN-zb&this work&4.2&5.2&0&0&3.4&0&4.9\\
&other calc.&3.81$^c$&4.01$^c$&0&0&0.05$^c$&0&3.94$^c$\\
\hline
GaN-wz&this work&4.5&4.4&4.4&5.4&0.024&0.22&4.5\\
&other calc.&3.74$^a$,4.54$^b$&3.67$^b$&3.19$^b$&4.54$^a$,5.4$^b$&0.58$^a$,0.49$^b$&0.48$^a$&4.3$^c$\\
&&4.88$^c$&3.69$^c$&3.75$^c$&6.54$^c$&0.49$^c$&&\\
\hline
InN-wz&this work&3.6&5.2&5.2&3.9&0.36&0.72&4.7\\
&other calc.&3.86$^a$,3.88$^b$&4.04$^b$&3.88$^b$&4.72$^a$,3.69$^b$&0.24$^a$,0.1$^b$&-0.08$^a$&3.92$^c$\\
&&3.66$^c$&3.51$^c$&4.11$^c$&4.26$^c$&0.15$^c$&&\\
\hline
\end{tabular}\\
$^a$ Ref. \cite{SLL10}.\\
$^b$ Ref. \cite{L07}.\\
$^c$ Ref. \cite{LMJ05}.\\
$^d$ Ref. \cite{KGM04}.\\

\end{flushleft}
\label{T_elastdP}
\end{table}

\begin{table}\footnotesize

\caption{The calculated band gaps in comparison with experimental values and other theoretical results for GaN and InN in WZ and ZB structures (all values in eV). In this work the calculations have been done within MBJLDA \cite{MBJLDA}, with the use of Abinit and Wien2k codes. In Abinit the $C_M$ parameter has been fitted to give experimental values.
}
\begin{flushleft}
\begin{tabular}{ccccc}
\hline
System&Exp.&Abinit &Wien2k&other calc. \\
\hline
GaN-zb&3.30$^j$&3.30 & 3.03 &3.05$^l$, 3.06$^l$\\
&&&&2.81$^m$,3.03$^n$\\
InN-zb&0.78$^j$&0.78 &0.64 &0.55$^l$, 0.63$^l$\\
&&&&\\
GaN-wz&3.50$^f$,3.51$^j$&3.50  &3.30 &3.56$^a$,3.47$^h$,3.50$^i$\\
&&&&3.23$^k$,3.26$^l$\\
&&&& 3.27$^l$, 3.21$^n$\\
InN-wz&0.65$^c$,0.63$^d$&0.65 &0.86 &0.60$^b$,0.69$^a$,0.65$^g$\\
&0.69$^e$&&&0.66$^k$,0.74$^l$\\
&&&& 0.63$^l$, 0.71$^n$\\
\hline
\end{tabular}\\

$^a$ Reference \cite{PRB_80_075202}\\
$^b$ Reference \cite{JApplPhys_104_013704}\\
$^c$ Reference \cite{PhysStatSolB_229_R1}\\
$^d$ Reference \cite{APL_83_4963}\\
$^e$ Reference \cite{JApplPhys_94_4457}\\
$^f$ Reference \cite{JApplPhys_83_1429}\\
$^g$ Reference \cite{JApplPhys_101_033123}\\
$^h$ Reference \cite{PRB_67_235205}\\
$^i$ Reference \cite{PRB_48_11810}\\
$^j$ Reference \cite{JApplPhys_94_3675}\\
$^k$ Reference \cite{MV10}\\
$^l$ Reference \cite{CK11}\\
$^m$ Reference \cite{MBJLDA}\\
\end{flushleft}
\label{T_Eg}
\end{table}

\section*{References}

%\bibliography{azotki}% Produces the bibliography via BibTeX.

\end{document}